\documentclass[preprint,sort&compress]{elsarticle}

\usepackage{lineno}
\usepackage[utf8]{inputenc}
\usepackage{graphicx}
\usepackage{lipsum}
\usepackage{amsmath}
\usepackage{txfonts}
\usepackage{amssymb}
\DeclareMathOperator\erf{erf}
\DeclareMathOperator\Dir{Dir}
\DeclareMathOperator\RMSE{{\it RMSE}}
\DeclareMathOperator\TPR{{\it TPR}}
\usepackage[para,online,flushleft]{threeparttable}
\usepackage[table]{xcolor}
\modulolinenumbers[10]

\journal{ArXiv}

\begin{document}
\begin{frontmatter}

\title{Bayesian analysis of data from segmented super-resolution images for quantifying protein clustering}

\author[add1,add2,fn2]{Tina Kos\v{u}ta}
\address[add1]{the Quantitative BioImaging lab, Facultat de Ci\`encies i Tecnologia, Universitat de Vic -- Universitat Central de Catalunya, Vic (Spain).}
\address[add2]{University of Ljubljana, Ljubljana (Slovenia).}
\fntext[fn2]{This research was performed while the author was a master student at the University of Ljubljana carrying out an Erasmus+ internship at the Universitat de Vic -- Universitat Central de Catalunya.}

\author[add1]{Marta Cullell-Dalmau}

\author[add3]{Francesca Cella Zanacchi}
\address[add3]{Nanoscopy and NIC@IIT, Istituto Italiano di Tecnologia, Genoa, (Italy).}

\author[add1]{Carlo Manzo\corref{corresp}}
\cortext[corresp]{Corresponding author}
\ead{carlo.manzo@uvic.cat}

\begin{abstract}
Super-resolution imaging techniques have largely improved our capabilities to visualize nanometric structures in biological systems. Their application further enables one to potentially quantitate relevant parameters to determine the molecular organization and stoichiometry in cells. However, the inherently stochastic nature of the fluorescence emission and labeling strategies imposes the use of dedicated methods to accurately measure these parameters. Here, we describe a Bayesian approach to precisely quantitate the relative abundance of molecular oligomers from segmented images. The distribution of proxies for the number of molecules in a cluster -- such as the number of localizations or the fluorescence intensity -- is fitted via a nested sampling algorithm to compare mixture models of increasing complexity and determine the optimal number of mixture components and their weights. We test the performance of the algorithm on {\it in silico} data as a function of the number of data points, threshold, and distribution shape. We compare these results to those obtained with other statistical methods, showing the improved performance of our approach. Our method provides a robust tool for model selection in fitting data extracted from fluorescence imaging, thus improving the precision of parameter determination. Importantly, the largest benefit of this method occurs for small-statistics or incomplete datasets, enabling accurate analysis at the single image level. We further present the results of its application to experimental data obtained from the super-resolution imaging of dynein in HeLa cells, confirming the presence of a mixed population of cytoplasmatic single motors and higher-order structures.
\end{abstract}

\begin{keyword}
Bayesian inference \sep Super-resolution microscopy \sep Protein clustering \sep Nested Sampling \sep Protein copy number
\end{keyword}

\end{frontmatter}

\newpage

\section*{Introduction}

The advent of super-resolution techniques has allowed the imaging of cellular structures beyond the limit imposed by diffraction, revealing unknown molecular arrangements at the nanoscale~\cite{Sahl2017,Sigal2018,Schermelleh2019}.  These techniques have rapidly evolved and their initial focus, i.e. resolving biological structures with unprecedented details, has shifted toward satisfying the demand of quantitative data to support specific biological hypotheses and models~\cite{Baddeley2018}. Along this line, an important effort has been devoted toward the development of calibrations and statistical methods for counting molecules in supramolecular arrangements~\cite{Annibale2011, Lee2012, Rollins2015, Nieuwenhuizen2015, Fricke2015, Nicovich2017}. As a matter of fact, the capability of counting molecules constitutes an important development to measure molecular stoichiometry, interactions, and organization, properties that play a fundamental role in signaling and other cellular functions~\cite{Yang2019}.\\
\begin{figure}[h!]
\centering
\includegraphics[width=.9\columnwidth]{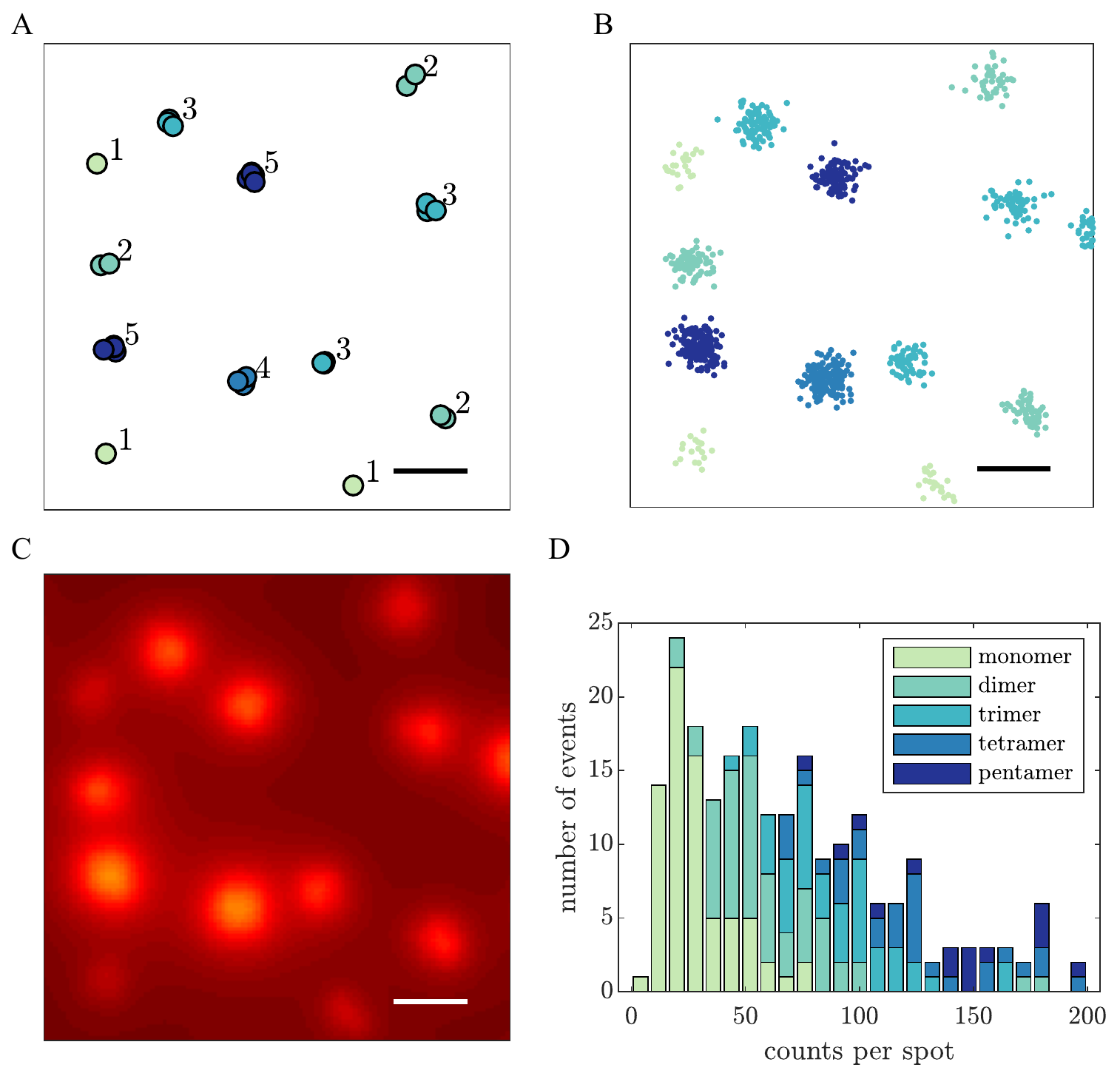}
\caption{Quantification of super-resolution imaging of protein nanoclusters. (A) A simulated image representing nanoclusters containing identical proteins with different stoichiometry. Super-resolution techniques such as SMLM, relying on fluorophores blinking, produce series of molecular localizations (B) that can be rendered into reconstructed images (C). Segmentation algorithms can be used to pinpoint spatially separated spots and extract the localization counts of each, displayed in the histogram of panel D. The stochastic nature of the process produces broad probability distributions from which is difficult to directly determine cluster stoichiometry. Scale bar: 100 nm.}
\label{fig:scheme}
\end{figure}
A ubiquitous features observed at the cell membrane, in the nucleus, and in organelles is the formation of protein clusters, i.e. little aggregates containing different copies of the same protein, either exclusively or in addition to other proteins~\cite{Garcia-Parajo2014,Feher2019}. Typically, these supramolecular arrangements of proteins occur at a scale of the order of tens of nanometers, thus comparable with the resolution provided by super-resolution techniques~\cite{Feher2019}. Therefore, it is not possible to distinguish or directly count each single protein within these nanostructures. At intermediate number densities ($\lesssim$ 500 $ \mu$m$^{-2}$), randomly distributed nanoclusters would mostly appear as resolvable features with size and shape comparable to the microscope point spread function (e.g., in STED nanoscopy), or as groups of molecular localizations (in single molecule localization microscopy, SMLM) with standard deviation comparable to the localization precision (Fig.~\ref{fig:scheme}). In these cases, manual and automatic feature-detection procedures~\cite{Chen2012, Manzo2014, Sage2015, Sage2019} and clustering algorithms~\cite{Mazouchi2015,Levet2015,Rubin2015,Griffie2016} have been devised to segment the image and automatically pinpoint spots/clusters, thus providing a quantification of the total fluorescence intensity per spot or of the number of localizations per cluster. These quantities are inherently stochastic as a consequence of the labeling strategy, the fluorophore photophysics, and the imaging protocol. As a consequence, chemically identical proteins produce a rather broad distribution of fluorescence intensities or number of localizations (Fig.~\ref{fig:scheme}D). However, these quantities generally conserve linearity (or at least positive correlation) with respect to the protein copy number. With the help of a reference, e.g. obtained by imaging and quantifying samples of known stoichiometry (tipically sparse monomeric proteins labeled as the actual sample), these quantities can be calibrated and used as proxies for protein counting~\cite{Ricci2015, Cella-Zanacchi2017, Martinez-Munoz2018}.\\
The output of this kind of calibration typically consists in a set of counts (e.g., corresponding to the number of localizations or to the fluorescence photon counts within a segmented region of the image) in a range $n=1,...,n_{max}$, from which the shape and parameters $\boldsymbol{\theta}$ describing the probability density function ({\it pdf}) associated to monomers $f_1(n|\boldsymbol{\theta})$ can be obtained. Under the assumption of linearity, this {\it pdf} allows to build the corresponding distribution for any oligomeric form of the protein, i.e. $f_i(n|\boldsymbol{\theta})$ with $i>1$. In fact, the rule of composition for the {\it pdf} of the sum of two independent random variables allows one to write:  
\begin{equation}
\label{eq:conv}
f_i(n|\boldsymbol{\theta}) = \sum_{m=1}^{\infty} f_1(m|\boldsymbol{\theta})f_{i-1}(n-m|\boldsymbol{\theta}) = f_1*f_{i-1}
\end{equation}
where the symbol $*$ indicates the convolution~\cite{Schmidt1996}.
The distribution of a population containing a mixture of $i$-mers with different weights $\alpha_i$, can thus be expressed as
\begin{equation}
\label{eq:lin_comb}
g(n, \boldsymbol{\alpha}|\boldsymbol{\theta}) = \sum_{i=1}^{K} \alpha_i f_i(n|\boldsymbol{\theta})
\end{equation}
where $K$ is the number of proteins forming the largest oligomer and $\sum_{i=1}^{K} \alpha_i =1$. 
Based on these formulas, it is in principle possible to obtain the weights corresponding to the different stoichiometries through the fit of the experimental distribution of the counts per spot $n$, as described earlier~\cite{Moertelmaier2005,Cella-Zanacchi2017}.\\
Although this procedure might seem rather straightforward, it presents some of the issues found for the analysis of finite mixtures, a thoroughly investigated topic in statistics~\cite{Mclachlan2000}, such as the presence of multiple maxima in the likelihood function and the need to know the number of components $K$. However, in contrast to finite mixtures, here we are dealing with non-identical {\it pdf}'s that are derived through the convolution and might not have a closed-form expression.\\
The fitting of a mixture model is generally performed via the optimization of the likelihood with either an expectation-maximization (EM) algorithm~\cite{Dempster1977}, or following a Bayesian approach based on Markov chain Monte Carlo (MCMC) methods~\cite{Mclachlan2000}. Assessing the exact number of component is an important and difficult task, since a precise determination of the free parameters $\alpha_i$ further relies on the knowledge of the maximum number of oligomers $K$ to consider.  This problem is generally solved by fitting a set of candidate models with different values of $K$ and determining the best based on same criterion. Since the likelihood increases by adding further parameters, the most popular criteria - such as the Bayesian information criterion (BIC)~\cite{Schwarz1978} and the Akaike's information criterion (AIC)~\cite{Akaike1973,Akaike1974} - introduce a penalty term for the number of parameters in the model to prevent overfitting. Both criteria can be derived in a Bayesian framework as approximated forms of the posterior probability using different prior probabilities~\cite{Burnham2004}. However, even though they are valid beyond the Bayesian context, they only provide correct estimations in the large-sample limit.\\
Tackling the problem in a Bayesian framework provides a systematic comparison of different models, while simultaneously providing the best associated parameters.  Not using the approximations of these information-criteria approaches enables a more accurate determination of the component weights, in particular when dealing with small datasets. The latter situation is often encountered in imaging experiments, when instrumental conditions induce changes of the calibration parameters $\boldsymbol{\theta}$ that do not allow the collection of large and uniform dataset, so that and the analysis must be performed at the single-image level.\\
In this article, we demonstrate how Bayesian inference can be efficiently applied to determine protein copy number from data obtained from segmented fluorescence images. We implement the Bayesian analysis using the nested sampling (NS) algorithm introduced by Skilling~\cite{Skilling2004, Skilling2006}. The algorithm provides an estimation of the Bayesian evidence to perform model ranking, calculates the model best-fit parameters and their confidence interval. We evaluate our method on synthetic data simulated in a wide range of conditions to take into account the effect of the amount of data, noise and threshold. We analyze its performance by calculating several statistical estimators and compare them with results obtained through conventional maximum likelihood estimation in combination with BIC and AIC. Our results indicate that our model ensures robust results,  outperforms traditional approaches for small dataset and offers good performance even with incomplete datasets, e.g. when the presence of spurious localizations impose the truncation of the data below a threshold. Its application to experimental data allows to confirm previous results from single-image analysis.\\
\section*{Results and Discussion}
\begin{figure}[hb!]
\centering
\includegraphics[width=.9\columnwidth]{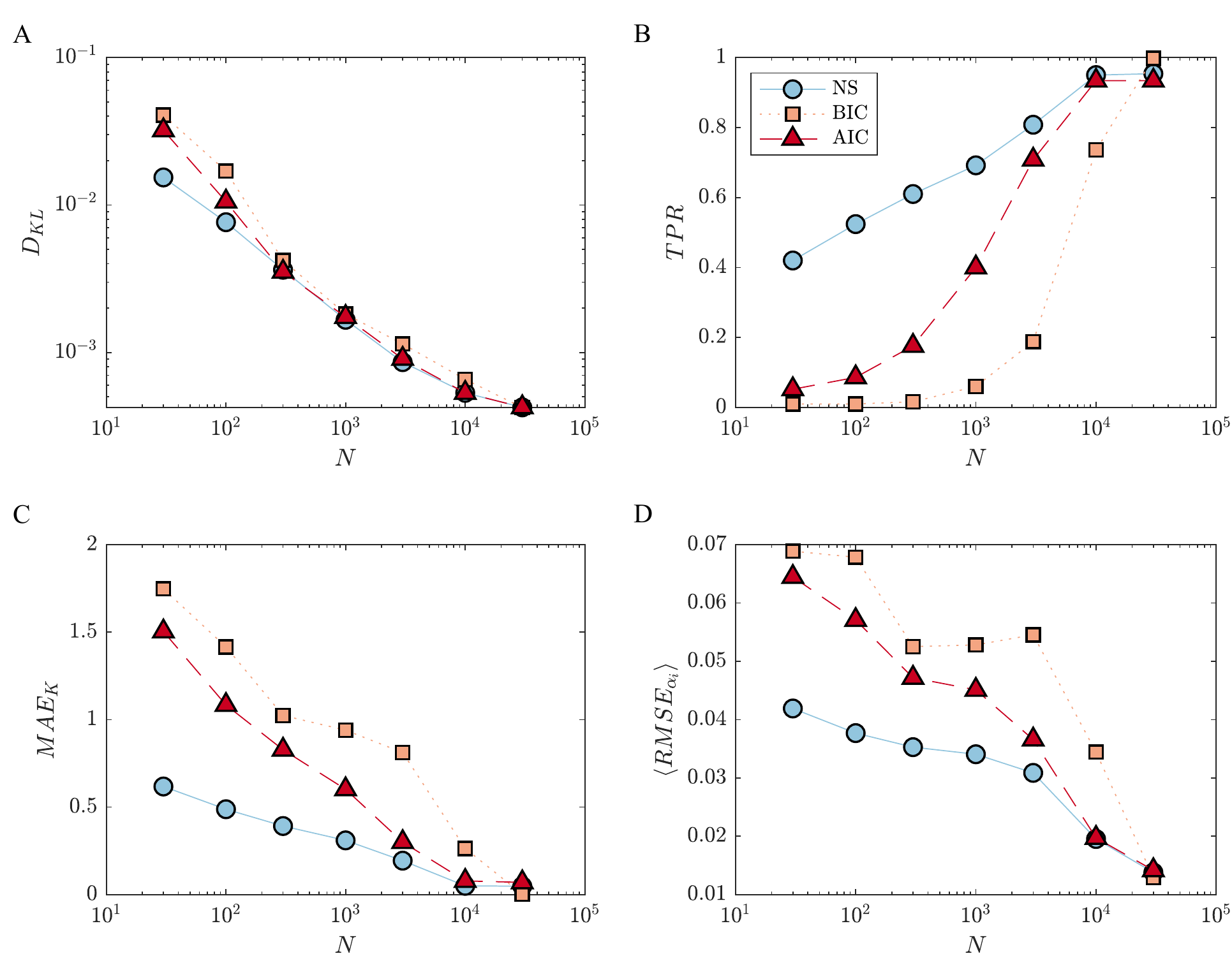}
\caption{Performance of the NS algorithm in comparison to BIC and AIC at varying the number of data points. (A) The Kullback-Leibler ($D_{KL}$) divergence. (B) The true positive rate $\TPR$. (C) The mean absolute error of the number of components $MAE_{K}$.  (D) The average root-mean square error of the weights $\langle \RMSE_{\alpha_{i}} \rangle$. Each point correspond to 500 simulations with parameters: $\boldsymbol{\alpha}=(0.11, 0.22, 0.33, 0.22, 0.11)$, $(\mu,\sigma)=(3.349,0.846)$ and analyzed with a Dirichlet prior with $\delta=1.5$.}
\label{fig:performance1}
\end{figure}

\begin{figure}[hbt!]
\centering
\includegraphics[width=1\columnwidth]{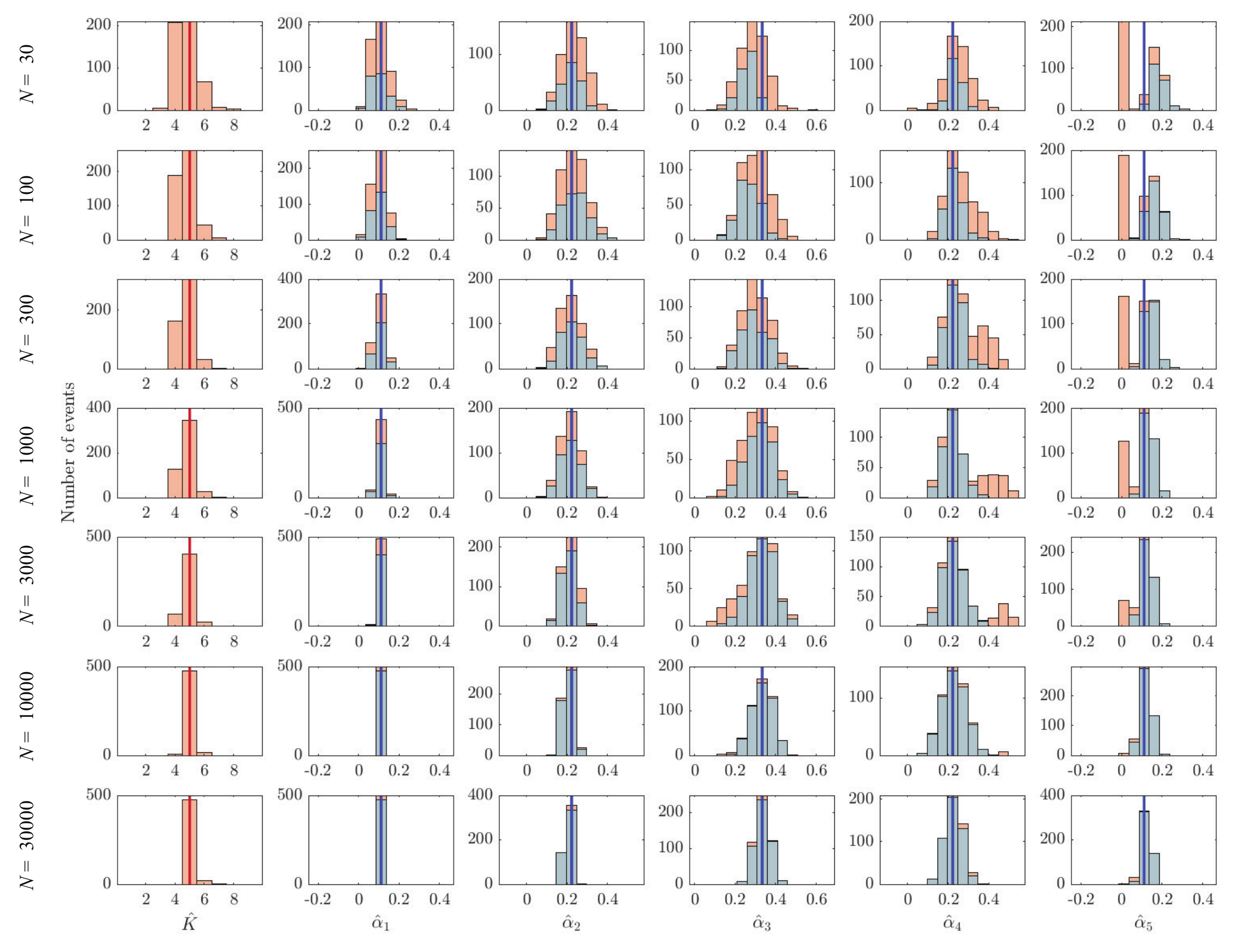}
\caption{Details of the performance of the NS algorithm at varying the number of data points. The first column shows the histogram of the value obtained for the estimator of the number of components $\hat{K}$ with respect to the ground truth value (red line). Columns from the second to the fifth represent the histograms of the value of the estimator of the component weight $\hat{\alpha}_i$ as obtained over all the simulation (pink) or over only on those for which the number of components has been correctly identified (light blue). The blue lines correspond to the ground truth values. Each line of plots corresponds to the results obtained for different number of data points $N$. Histograms correspond to 500 simulations with parameters: $\boldsymbol{\alpha}=(0.11, 0.22, 0.33, 0.22, 0.11)$, $(\mu,\sigma)=(3.349,0.846)$ and analyzed with a Dirichlet prior with $\delta=1.5$.}
\label{fig:performance2}
\end{figure}

\begin{figure}[hbt!]
\centering
\includegraphics[width=.9\columnwidth]{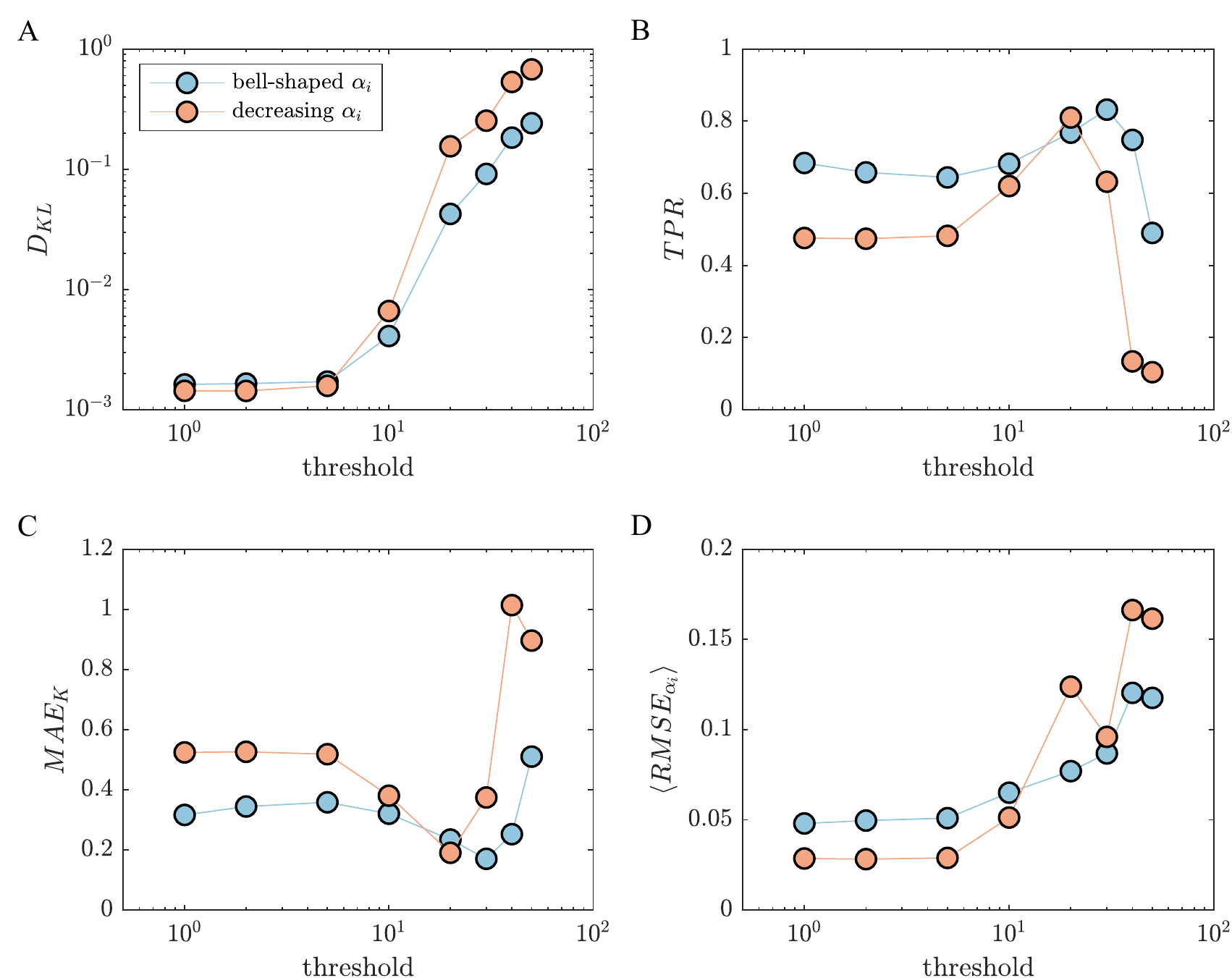}
\caption{Performance of the NS algorithm for different threshold values and different distribution of weights. (A) The Kullback-Leibler ($D_{KL}$) divergence. (B) The true positive rate $\TPR$. (C) The mean absolute error of the number of components $MAE_{K}$. (D) The average root-mean square error of the weights $\langle \RMSE_{\alpha_{i}} \rangle$. Each point correspond to 500 simulations with parameters: $N=1000$, $\boldsymbol{\alpha}=(0.11, 0.22, 0.33, 0.22, 0.11)$ (bell-shaped) or $\boldsymbol{\alpha}=(0.33, 0.27, 0.20, 0.13, 0.07)$ (decreasing), $(\mu,\sigma)=(3.349,0.846)$ and analyzed with a Dirichlet prior with $\delta=1.5$.}
\label{fig:performanceth}
\end{figure}

\paragraph{Bayesian inference and model selection}
We consider a dataset $\{x\}$ composed of $N$ measurements with discrete values $n=1,...,n_{max}$ corresponding to a process described by the {\it pdf} given in Eq.~\ref{eq:lin_comb}. Our objective is estimating the component weights $\boldsymbol{\alpha}$ for a specific model including at most $K=K'$ oligomeric species. In a Bayesian inference framework~\cite{Hines2015}, we need to calculate the posterior distribution which is expressed according to Bayes' theorem as
\begin{equation}
\label{eq:bayes_inf}
P(\boldsymbol{\alpha}_{K'}|K',x)=\frac{P(x|\boldsymbol{\alpha}_{K'},K') P(\boldsymbol{\alpha}_{K'}|K')}{P(x|K')}=\frac{\mathcal{L}(\boldsymbol{\alpha}_{K'}) \pi(\boldsymbol{\alpha}_{K'})}{\mathcal{Z}_{K'}},
\end{equation}
where $P(n|\boldsymbol{\alpha}_{K'},K')=\mathcal{L}(\boldsymbol{\alpha}_{K'})$ represents the likelihood,  $P(\boldsymbol{\alpha}_{K'}|K')=\pi(\boldsymbol{\alpha}_{K'})$ the prior probability and $P(n|K')=\mathcal{Z}_{K'}$ is the evidence. The right-hand side symbols refer to the notation used in the classical literature~\cite{Skilling2006}. The evidence is generally difficult to calculate, since it requires integration over all possible parameter values. In most of the cases, it is sufficient to estimate the posterior up to a multiplicative constant and the evidence is thus ignored.\\ 
However, if we want to statistically compare and rank several models characterized by different $K_j$, we need to apply again the Bayes' theorem as:
\begin{equation}
\label{eq:bayes_mod}
P(K_j|x)=\frac{P(x|K_j) P(K_j)}{P(x)}.
\end{equation}
In Eq.~\ref{eq:bayes_mod}, the term $P(x|K_j)$, corresponding to the evidence in Eq.~\ref{eq:bayes_inf}, takes the meaning of a likelihood. As such, it is crucial for the evaluation of the probability of a model. In fact, assuming that all the models have the same prior probability $P(K_j)$, and considering that $P(x)$ is a constant term that only depends on the data, the model posterior can be approximated as:
\begin{equation}
\label{eq:bayes_mod}
P(K_j|x) \propto P(x|K_j) =\mathcal{Z}_{K_j},
\end{equation}
and thus used to rank the different models.
\paragraph{The nested sampling algorithm} 
As mentioned above, the estimation of $\mathcal{Z}_{K_j}$ involves the calculation of the likelihood $\mathcal{L}(\boldsymbol{\alpha}_{K_j})$, and the solution of the integral
\begin{equation}
\label{eq:Zintegral}
\mathcal{Z}_{K_j}=\int \mathcal{L}(\boldsymbol{\alpha}_{K_j}) \pi(\boldsymbol{\alpha}_{K_j}) d\boldsymbol{\alpha}_{K_j},
\end{equation}
that become complicated when $\boldsymbol{\alpha}_{K_j}$ has more than a few dimensions. 
The likelihood can be easily calculated considering the number of events $h$ recorded for each value $n$ of the discrete variable $x$:
\begin{equation}
\label{eq:likeli}
\mathcal{L}(\boldsymbol{\alpha}_{K_j})=\prod_{n}   g(n,\boldsymbol{\alpha}_{K_j}) ^{h(n)}.
\end{equation}
For the prior, we use the symmetric Dirichlet function~\cite{Jasra2005} 
\begin{equation}
\label{eq:Dirichlet}
\Dir(\boldsymbol{\alpha}_{K_j}|\delta)=\frac{ \Gamma(\delta K_j)} {\Gamma (\delta)^{K_j} }\prod_{i=1}^{K_j} \alpha_{K_j,i}^{\delta-1},
\end{equation}
for different values of $\delta$. In the following, for the sake of simplicity, we will use only the subscript $j$ instead of $K_j$ to indicate a specific model. \\
For the calculation of the evidence, we use the NS approach~\cite{Skilling2004, Skilling2006}, that allows us to reduce the integral in Eq.~\ref{eq:Zintegral} to one dimension:
\begin{equation}
\label{eq:ZintegralNS}
\mathcal{Z}_{j}=\int_0^1 \mathcal{L}(X_j) dX_j,
\end{equation}
where $\mathcal{L}(X_j)$ is the inverse of the so-called ``prior mass''~\cite{Skilling2006}
\begin{equation}
\label{eq:XintegralNS}
X(\lambda)=\int_{\mathcal{L}(\boldsymbol{\alpha})>\lambda} \pi(\boldsymbol{\alpha}) d\boldsymbol{\alpha}.
\end{equation}
In practice, the evidence in Eq.~\ref{eq:ZintegralNS} is calculated as a sum
\begin{equation}
\label{eq:evid_sum}
\mathcal{Z}_{j} \approx \sum_{k} \mathcal{L}_k(\boldsymbol{\alpha}_j) \varw_{k}
\end{equation}
over a number of ``particles'' corresponding to values of the vector of parameters $\boldsymbol{\alpha}$ and with quadrature weights $\varw_{k}=\Delta X_k$. For this calculation, we follow a NS implementation similar to the one recently described for the analysis of single-particle trajectories~\cite{Thapa2018}. Namely, we first generate a random set of particles and sort them by their likelihood.  We select the one with the minimum likelihood and assign the weight $\varw_{k=1} = \frac{1}{N+1}$, where $N$ is the number of particles, to calculate the first term of the sum in Eq.~\ref{eq:evid_sum}. The particle is thus removed from the set. A new particle is generated by copying one of the surviving particles and performing a random walk by the Metropolis-Hasting algorithm. The particle is moved in a random direction with a step length drawn from a normal distribution with a mean equal to zero and a standard deviation initially set to $0.1$ and updated at every iteration to ensure an acceptance rate of $\sim$50\%~\cite{Sivia2006book,Feroz2008}.
The movement is constrained in the domain $[0,1]$ through periodic boundary conditions. After each iteration, all the coordinates are normalized so to verify the condition $\sum_{i} \alpha_i =1$. The Metropolis-Hasting algorithm is conditioned such to have a likelihood higher than the one of the removed particle and with an acceptance ratio based on the particle prior~\cite{Feroz2008}. After each run, a new term is added to the sum in Eq.~\ref{eq:evid_sum}, consisting in the product of the minimum likelihood and the weight $\varw_{k} =\varw_{k-1} \frac{N}{N+1}$, to take into account of the reduction of the range of the ``prior mass'' at each iteration~\cite{Skilling2006,Thapa2018}. The procedure is repeated until the ratio $ \frac{\mathcal{Z}_{res}}{\mathcal{Z}_j}<10^{-5}$~\cite{Thapa2018}, where $\mathcal{Z}_{res}=\varw_{k} \sum_{m} \mathcal{L}_m$ represents the residual evidence after $k$ steps, with the sum running over the surviving particles. 
\begin{figure}[hbt!]
\centering
\includegraphics[width=.9\columnwidth]{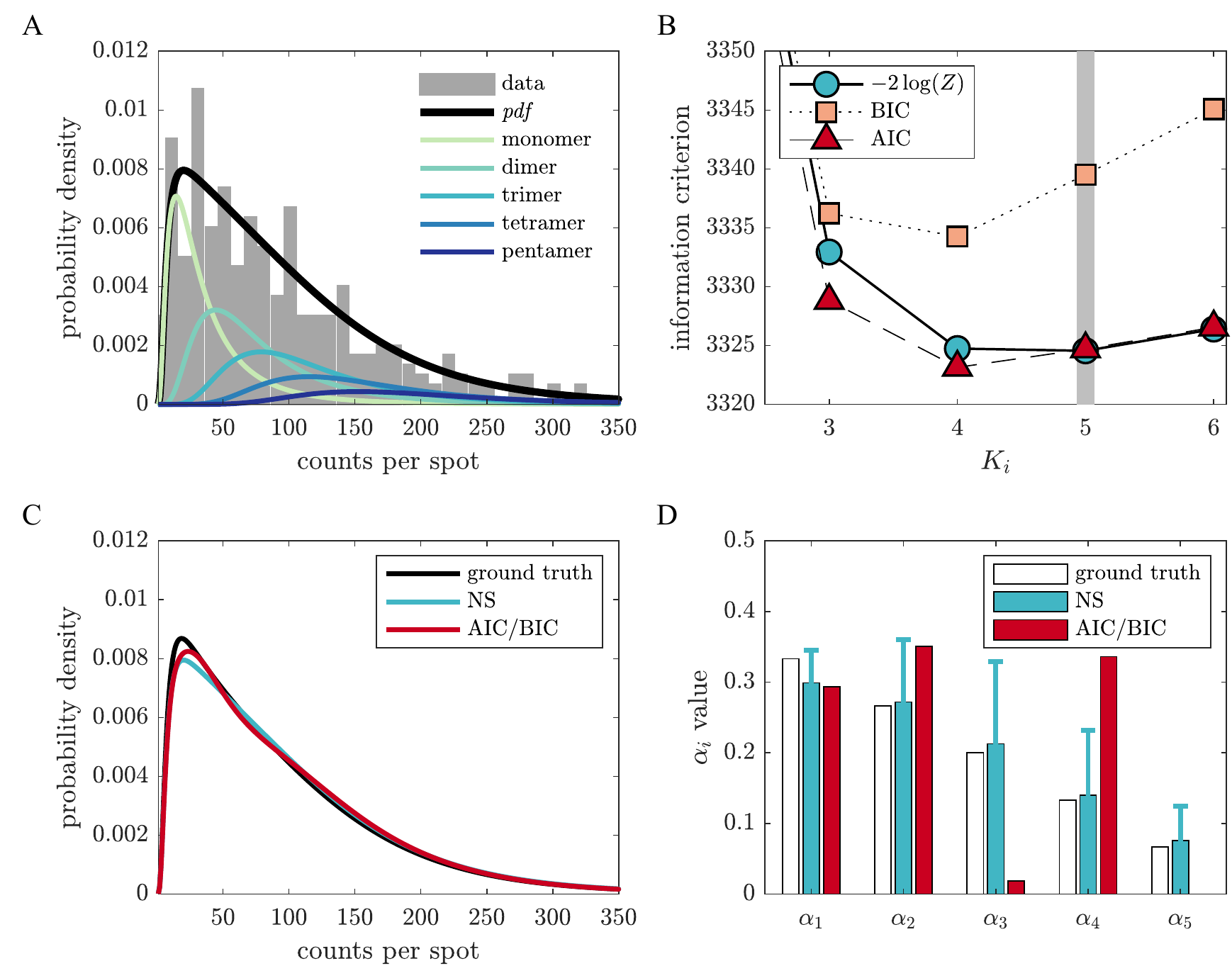}
\caption{Results of the application of the NS algorithm to a simulated dataset. (A) Histogram of the simulated data and results of the fit. The color curves correspond to the {\it pdf}'s of different species multiplied by the corresponding weight. (B) Information criteria at varying the number of components considered for the inference. The NS shows a minimum corresponding to the ground truth value $K_{gt}$. (C) Bar plot of the  weights determined by the different approaches. The NS retrieves weights compatible with ground truth values within the uncertainty, even for small datasets.The dataset was simulated with parameters: $N=300$, $\boldsymbol{\alpha}=(0.33, 0.27, 0.20, 0.13, 0.07)$, $(\mu,\sigma)=(3.349,0.846)$ and analyzed with a Dirichlet prior with $\delta=1.0$.}
\label{fig:results1}
\end{figure}
\paragraph{Performance on synthetic data}
We tested the algorithm on synthetic data simulated by varying several parameters for mimicking different experimental conditions.
For space limitations, here we only report results obtained in a few relevant cases compared with the ground truth ({\it gt}). In Fig.~\ref{fig:results1}, we show as an example the detailed results of the analysis of a single simulated dataset composed by $300$ datapoints and with $K_{gt}=5$ components with decreasing weights. The histogram of the data is shown in Fig.~\ref{fig:results1}A, together with the generating distribution and the {\it pdf}'s of the different oligomeric species, each multiplied by the corresponding ground truth weight. The algorithm runs the NS analysis iteratively for increasing values of $K_{j}$ and evaluates $\mathcal{Z}$ (as well as the BIC and the AIC), stopping when a maximum in $\mathcal{Z}$ is detected. The results of the inference are reported in Table~\ref{tab:results1} and in Fig.~\ref{fig:results1}B, where to facilitate the comparison with the other criteria we plot $-2 \log \mathcal{Z}$. Although the different methods retrieve best-fit curves close to the {\it gt} (Fig.~\ref{fig:results1}C), the NS finds the right value of components whereas both the BIC and the AIC show minima at smaller values. Moreover, the NS recovers weights that are compatible with the {\it gt} values (Fig.~\ref{fig:results1}D and Table~\ref{tab:results1}). We find this result quite remarkable considering the large overlap between the {\it pdf}'s of consecutive components, the small dataset and the low value of $\alpha_5$.  In fact, although in this case we have only $20$ datapoints associated to pentameric clusters, the algorithm is still able to determine the weight of this species rather precisely.\\
\begin{table}[h!]
\caption{Results of Bayesian inference on a simulated dataset of $300$ points with $K_{gt}=5$ and decreasing $\alpha$'s (Fig.~\ref{fig:results1}).}
\label{tab:results1}
\centering\resizebox{\textwidth}{!}{
\begin{tabular}{c c c c c c c c | c c }
\hline
$K_j$ & $\log(\mathcal{Z}_j)$ & $\alpha_1$ & $\alpha_2$ & $\alpha_3$ & $\alpha_4$ & $\alpha_5$ & $\alpha_6$ & BIC & AIC   \\\hline
2 & -1693.6$\pm$0.3 & 0.21$\pm$0.04 & 0.79$\pm$0.04 & -- & -- & -- & -- &3387.7  & 3384.0 \\
3 & -1666.5$\pm$0.3 & 0.32$\pm$0.05 & 0.13$\pm$0.09 & 0.54$\pm$0.07 & -- & -- & -- & 3336.2 &  3328.8  \\
4 & -1662.4$\pm$0.3 & 0.30$\pm$0.05 & 0.26$\pm$0.10 & 0.19$\pm$0.13 & 0.25$\pm$0.09 & -- & -- & 3334.3  & 3323.1 \\
\rowcolor{gray!30} 
5 & -1662.3$\pm$0.3 &0.30$\pm$0.05 & 0.27$\pm$0.09 & 0.21$\pm$0.12 & 0.14$\pm$0.09 & 0.08$\pm$0.05 & -- & \cellcolor{white} 3339.5  &\cellcolor{white} 3324.7   \\
6 & -1663.2$\pm$0.3 &0.29$\pm$0.05 & 0.30$\pm$0.11 & 0.19$\pm$0.12 & 0.13$\pm$0.07 & 0.05$\pm$0.04 & 0.03$\pm$0.03 & 3345.1 &  3326.6 \\
\hline
$K_{gt}=5$ & -- & 0.33 & 0.27 & 0.2 & 0.13 & 0.07 & --  &  -- & --  \\
\hline
\end{tabular}
}
\end{table}
To further test the performance of the method, we simulated data according to the same model presented above at varying the number of data points. We run $500$ simulations in each case and quantify the overall performance by evaluating the goodness of the fit by the Kullback-Leibler ($D_{KL}$) divergence, the true positive rate ($\TPR$, i.e. the fraction of cases for which the estimated number of components $\hat{K}$ was equal to $K_{gt}$), the mean absolute error ($MAE$) for the number of components $MAE_{K} = \langle |K_{gt}-\hat{K}| \rangle$, and the root-mean square error ($\RMSE$) of the weights $\RMSE_{\alpha_{i}} = \sqrt{ \langle  (\alpha_{gt,i}-\hat{\alpha}_{i})^2\rangle}$.\\ 
The values obtained for a specific set of parameters are shown in Fig.~\ref{fig:performance1}. First, we calculated the $D_{KL}$ of the fitted distributions $q(n)$ with respect to the ground truth distribution $p(n)$ as $D_{KL} = \sum_{n=1}^{\infty} p(n) \log\frac{p(n)}{q(n)} $ as a measure of the information gain (Fig.~\ref{fig:performance1}A). All the methods show similar performance at reproducing the ground truth distribution, with an expected improvement as a function the number of data points.  
Moreover, for small $N$, Fig.~\ref{fig:performance1}A indicates that the NS method provides  better fits as compared to the other methods. However, $D_{KL}$ only tells us about the goodness of the fit, but does not report about the precision in the determination of the model or in the value retrieved for free parameters. The plot of the $\TPR$ (Fig.~\ref{fig:performance1}B) further shows that the NS outperforms the other methods in detecting the right number of components for small datasets and thus rules out the possibility of overfitting. Even if the values obtained for the $\TPR$ might seem low ($~0.6-0.7$ for $N=300-1000$), they are 2 to 3 times better than the other methods. In addition, one must consider that even when not retrieving the exact value of components, the Bayesian approach estimates $K$ with a very small error. This can be appreciated by the plot in Fig.~\ref{fig:performance1}C, where for the NS we obtain values of the mean absolute error smaller than $1$ and in more detail from the histograms of Fig.~\ref{fig:performance2}. Furthermore, we also measured the deviation of the estimated values of $\alpha_{i}$ with respect to the ground truth. In Fig.~\ref{fig:performance2} we report the histograms for each component weight, whereas for simplicity in Fig.~\ref{fig:performance1}D we only show the average of $\RMSE$ over all the weights $\langle \RMSE_{\alpha_{i}} \rangle $.  These results confirm that the parameters calculated by the NS better reproduce the true weights of the oligomeric components, as also observed in Fig.~\ref{fig:results1}D.\\
Typically, in datasets obtained from segmented fluorescence images, the number of events corresponding to low counts is overestimated as a consequence the presence of spurious spots/localizations resulting from noise and/or false positive detection. To filter out these data, a simple strategy consists in cutting down the dataset by removing events below a positive threshold. We explored the effect of this left truncation on the results of the Bayesian inference, as reported in Fig.~\ref{fig:performanceth}. For distributions with both decreasing and bell-shaped weights $\alpha_i$ a moderate threshold improves the performance of the analysis in determining the true number of components.  This effect might seem rather counter-intuitive at first glance, but can be interpreted as the consequence of a reduction in the effective weight of the populations with low stoichiometry (monomer, dimer) with respect to those represented in the right tail of the distribution, thus favouring fits with more components. In fact, it is accompanied by an increase of the mean error on the determination of the parameters $\alpha_i$ (Fig.~\ref{fig:performanceth}), mostly driven by the loss of information on the low-stoichiometry components. This reasoning is further supported by the comparison of the results for different distributions of weights, showing a minor effect on bell-shaped $\alpha_i$ with respect to decreasing $\alpha_i$ as the result of the different value of the weight corresponding to the largest oligomeric species. In both cases, as the threshold is further increased, the overall loss of information generates an expected decrease of the performance. We performed further simulations and analysis at changing the number of ground truth components $K_{gt}$, the distribution of weights $\boldsymbol{\alpha}_{gt}$, the parameters defining the calibration function ($\mu,\sigma$) and the prior probability $\delta$. 
%

\begin{figure}[hbt!]
\centering
\includegraphics[width=.9\columnwidth]{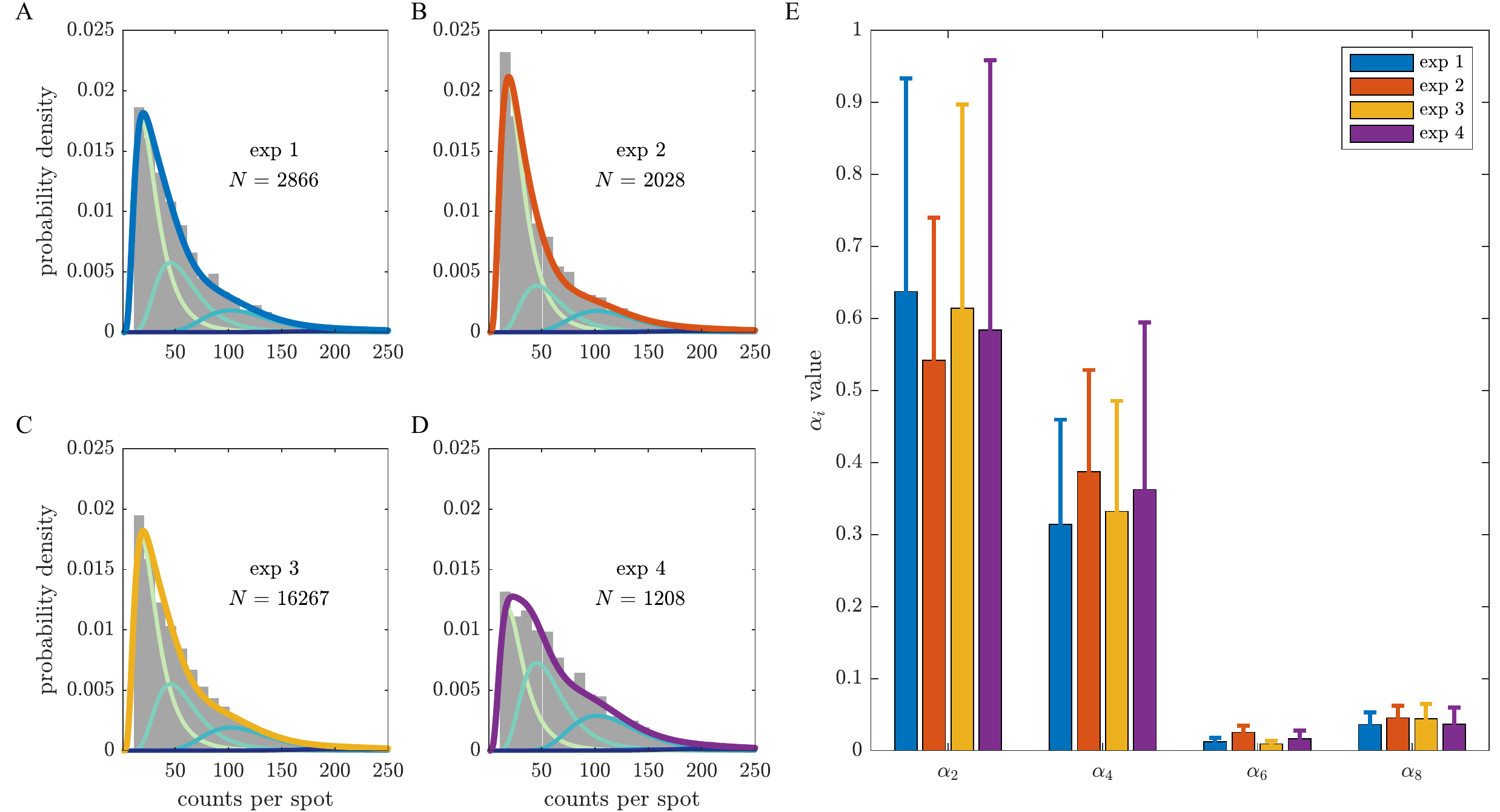}
\caption{Results of the application of the NS algorithm to STORM data. (A-D) Histograms of the data obtained from 4 different STORM image and results of the fits. The color curves correspond to the {\it pdf}'s of different species multiplied by the corresponding weight.  (E) Bar plot of the  weights determined by the NS approach on the 4 datasets. The weights of the dimeric and clustered species show compatible values. The data were analyzed including the {\it pdf}'s corresponding to monomers, dimers and clusters of dimers with $(\mu,\sigma)=(3.227,0.569)$, and a Dirichlet prior with $\delta=1.5$.}
\label{fig:exp_data}
\end{figure}

\paragraph{Analysis of STORM data}
We further test our method on datasets obtained from four individual STORM images of dynein intermediate chain (DIC) fused to GFP stably expressed in HeLa cells~\cite{Cella-Zanacchi2019}. For these imaging conditions, a calibration function was obtained following an approach based on DNA-origami~\cite{Cella-Zanacchi2017,Cella-Zanacchi2019}. The super-resolution images of dynein labeled with anti-GFP primary and secondary antibodies revealed groups of localizations that were segmented via a clustering algorithm. Events corresponding to less than 10 counts per spot were not taken into account. The data were fitted via the NS algorithm, as described for the simulations using the as a prior the Dirichlet function with $\delta=1.5$ (Fig.~\ref{fig:exp_data}). As discussed in an earlier work, dynein is a homodimer containing two copies of DIC, but DIC may not be incorporated into a fully-assembled functional dynein motor complex and may exist as a monomer in cells~\cite{Cella-Zanacchi2017}. Therefore, we included in the fit the {\it pdf}'s corresponding to monomers, dimers and clusters of dimers, thus considering as negligible the occurrence of mixtures of dimers and monomers. The NS algorithm consistently finds that all the experiments are best fitted with $K=5$ components, whereas the BIC and the AIC find different values from 6 to 9. Moreover, out of the variable (35-60\%) fraction of monomeric DIC, the NS shows that the functional dynein motor complex has a highly consistent distribution, corresponding to a $\sim60\%$ of dimers, $\sim35\%$ of tetramers and the rest of clusters containing 3 and 4 dimers (Fig.~\ref{fig:exp_data}D).

\section*{Conclusions}
We introduced a Bayesian approach based on the nested sampling algorithm for the multicomponent fitting of data obtained through the segmentation of super-resolution image. As shown by the application of the method to synthetic data, the algorithm assess the correct number of components and determine the component weights with higher precision as compared to BIC and AIC, thus allowing one to accurately estimate cluster stoichiometry in biological samples. Moreover, the Bayesian analysis allows one to use physical information by defining proper parameter ranges through the priors.\\
Notably, the method offers largest improvements over traditional approaches for small dataset. Therefore, it enables robust single-image quantification, eliminating the need for cumulating data obtained from different images that can introduce variability as a result of different experimental and calibration conditions. Furthermore, it also performs rather well with incomplete, left-truncated datasets. This situation is typically encountered when dealing with real data, where a threshold is introduced to filter out the overcount of events associated to spurious spots/localizations resulting from noise and/or false positive detection.\\
Although we presented examples from SMLM, the method can be equally applied on data extracted from stochastic or deterministic super-resolution techniques as long as the calibration function corresponding to an oligomeric species can be estimated. Despite the fact that in this work we assume that the parameters characterizing the shape of the {\it pdf} are previously known from calibration measurements, our method can be generalized to allow their inference from the Bayesian analysis with a proper choice of the corresponding priors~\cite{Dose2003, Sivia2006book}. Moreover, for the case of SMLM data, the method can be further extended to also include the Bayesian treatment of spatial and temporal information and thus bypass the previous segmentation step.\\
Our method improves the post-processing step present in a type of workflows used to estimate and quantify molecular clustering in cells from super-resolution images. Molecular clustering is thought to have a fundamental role in cell signalling, therefore the ability to detect it and precisely quantify is necessary to reveal the biological implications of molecular self-organization. Although several quantitative approaches - also involving super-resolution techniques - have been proposed, several questions are still open. Recently, the existence of nanoclusters in T-cells has been questioned on the basis of contrasting results obtained with SMLM techniques~\cite{Rossboth2018,Feher2019}. Therefore, improvements in hardware, probes, and pre- and post-processing software are necessary to obtain solid quantitative information about these biological processes. In this scenario, we hope that our method can help to better quantify super-resolution data and thus contribute to determine the biological relevance of clustering.

\section*{Materials and Methods}
\paragraph{Software implementation and simulations} 
The code for implementing the NS algorithm was written in both Matlab (The MathWorks, Inc., Natick, Massachusetts, United States) and R (R Foundation for Statistical Computing, Vienna, Austria).\\
The NS algorithm used for this study was implemented on 30 particles with a depth of 40 iterations. For the Metropolis-Hastings, the random motion of a particle is performed by adding a variable step in a random direction. To ensure an acceptance rate of $\sim$50\%, the step length is drawn from a normal distribution with standard deviation $\sigma_{step}$ initially set to $0.1$ and updated at each iteration as~\cite{Sivia2006book,Feroz2008}
\begin{equation}
\sigma_{step}=\left \{ \begin{array}{rl}  \sigma_{step} \cdot  e^{1/A} & \text{if } A>R \\
							  \sigma_{step}  \cdot e^{-1/R} & \text{if } A \leq R \\
								\end{array} \right. ,
\end{equation}
where $A$ and $R$ are the numbers of accepted and rejected samples, respectively.
We explored the behavior of the algorithm for different prior probabilities corresponding to the Dirichlet distribution with $\delta=1$ (uniform distribution), $\delta=0.5$ and $\delta=1.5$. \\
Data corresponding to simulations of localization counts obtained from monomeric clusters were generated considering a discretized lognormal {\it pdf}
\begin{equation}
\label{eq:lognorm}
f_1(n|\mu,\sigma) = \frac{1}{2}   \bigg[ \erf \bigg(\frac{\mu-\log(n-1)}{\sqrt{2}\sigma } \bigg) - \erf \bigg(  \frac{\mu-\log(n)}{\sqrt{2}\sigma } \bigg) \bigg] ,
\end{equation}
with parameters $(\mu, \sigma)$ equal to $(3.349, 0.846)$ and $(3.227, 0.569)$, since they have been recently shown to accurately approximate the output of STORM imaging in different experimental conditions~\cite{Cella-Zanacchi2017, Cella-Zanacchi2019}. Localization counts in oligomeric clusters of $m$ proteins were obtained by the sum of $m$ random numbers obtained as described above. 
%
\paragraph{Sample preparation for STORM microscopy} HeLa IC74-mfGFP stably transfected cell line (from Takashi Murayama lab, Department of Pharmacology, Juntendo University School of Medicine, Tokyo, Japan) were plated on 8-well Lab-tek 1 coverglass chamber (Nunc) and grown under standard conditions (DMEM, high glucose, pyruvate (Invitrogen 41966052) supplemented with 10\% FBS, 2 mM glutamine and selected with 400 $\mu$g/mL Hygromycin).  Cells were fixed with PFA (3\% in PBS) at RT for 7 minutes. Cells were then incubated at RT with blocking buffer (3\% (wt/vol) BSA (Sigma) in PBS and 0.2\% Tryton. In HeLa IC74-mfGFP stably transfected cells, dynein intermediate chain-green fluorescent protein (GFP) was immuno-stained with primary antibody (chicken polyclonal anti GFP, Abcam 13970) diluted 1:2000 in blocking buffer for 45 minutes at room temperature. Cells were rinsed 3 times in blocking buffer for 5 minutes and incubated for 45 minutes with secondary antibodies donkey-anti chicken labeled with photoactivatable dye pairs for STORM (Alexa Fluor 405-Alexa Fluor 647). 

\paragraph{STORM microscopy}
Imaging was performed with an oil immersion objective (Nikon, CFI Apo TIRF 100x, NA 1.49, Oil), repeated cycles of activation (405 nm laser), and readout (647 nm laser) using TIRF illumination. During experiments the focus was locked through the Perfect Focus System (Nikon) and imaging was performed on an EMCCD camera (Andor iXon X3 DU-897, Andor Technologies).
A commercial N-STORM microscope (Nikon Instruments) was used to acquire 40,000 frames at a 33 Hz frame rate. An excitation intensity of $\sim$0.9 kW/cm$^2$ for the 647 nm readout laser (300 mW MPB Communications, Canada) and an activation intensity of $\sim$35 W/cm$^2$ for the 405 nm activation laser (100 mW, Cube Coherent, CA) were used.  STORM imaging buffer was used containing GLOX solution as oxygen scavenging system (40 mg/mL Catalase [Sigma], 0.5 mg/mL glucose oxidase, 10\% glucose in PBS) and MEA 10 mM (Cysteamine MEA [SigmaAldrich, \#30070-50G] in 360 mM Tris-HCl). 

\paragraph{STORM data analysis} Localization and reconstruction of STORM images were performed using custom software (Insight3, kindly provided by Bo Huang, University of California) by Gaussian fitting of the single molecules images to obtain the localization coordinates. The final image is obtained plotting each identified molecule as a Gaussian spot with a width corresponding to the localization precision (10 nm) and corrected for drift. A custom code implementing a distance-based clustering algorithm, was used to identify spatial clusters of localizations. The localizations list was first binned to 20 nm pixel size images that were filtered with a square kernel ($7 \times 7$ pixel$^2$) and thresholded to obtain a binary image. Only the localizations lying on adjacent (6-connected neighbours) non-zero pixels of the binary image were considered for further analysis. To select the sparse dynein contribution large clusters were filtered out setting a threshold on the maximum number of localizations (1000 localizations/cluster). 

\section*{Author Contributions}
C.M. designed and supervised the research. T.K. and C.M. implemented the algorithm, performed the simulations and analyzed the synthetic data. M.C.-D. and C.M. analyzed the experimental data. F.C.-Z. performed live-cell imaging. C.M. wrote the article. All authors discussed the results, read and approved the manuscript.

\section*{Acknowledgments}
HeLa IC74 cell line was a kind gift of Takashi Murayama, Department of Pharmacology, Juntendo University School of Medicine, Tokyo, Japan. We thank Dr. Angel Sandoval Alvarez, ICFO, Barcelona  for helping with cell maintenance and transfections. 
C.M. acknowledges funding from FEDER/Ministerio de Ciencia, Innovaci\'{o}n y Universidades -- Agencia Estatal de Investigaci\'{o}n through the ``Ram\'{o}n y Cajal'' program 2015 (Grant No. RYC-2015-17896), and the ``Programa Estatal de I+D+i Orientada a los Retos de la Sociedad'' (Grant No. BFU2017-85693-R); from the Generalitat de Catalunya (AGAUR Grant No. 2017SGR940). C.M. also acknowledges the support of NVIDIA Corporation with the donation of the Titan Xp GPU. C.M. and M.C.-D. acknowledge funding from the PO FEDER of Catalonia 2014-2020 (project PECT Osona Transformaci\'{o} Social, Ref. 001-P-000382). T.K. acknowledges the support of the Erasmus+ program of the European Union. F.C.-Z. acknowledges the Nikon Imaging Center at the Italian Institute of technology.

\bibliography{main}

\end{document}